**Who is the author? A legal and normative view of authorship in Generative AI-aided academic works**

**David M. Pereira**

REQUIMTE/LAQV, Laboratory of Pharmacognosy, Department of Chemistry, Faculty of Pharmacy, University of Porto, R. Jorge Viterbo Ferreira, 228, 4050-313 Porto, Portugal

**Abstract**

The widespread adoption of generative artificial intelligence (GenAI) tools in higher education has fundamentally altered the conditions under which academic work is produced, challenging long-standing assumptions about authorship, responsibility, and learning. While much of the existing literature has focused on technical, ethical, or pedagogical implications of GenAI, comparatively little attention has been paid to the legal and normative aspects of authorship in AI-aided academic work.

In this work, we examine how the use of GenAI intersects with the concept of authorship as understood within European regulatory and institutional frameworks.

Drawing primarily on European copyright law, notably the requirement of human intellectual creation, the paper argues that authorship functions as a qualitative threshold rather than a binary attribute. Authorship may remain attributable to the student where GenAI operates as cognitive support under human intellectual control. By contrast, attribution becomes legally and normatively disputable once AI output displaces creative autonomy.

The analysis places this doctrinal framework alongside broader regulatory principles arising from the AI Act, data protection law, and emerging suprainstitutional governance practices in higher education.

We propose a qualitative threshold framework designed to assist in authorship-sensitive assessment of GenAI-aided academic work. This framework provides criteria for distinguishing legitimate AI-assisted academic production from practices that undermine authorship, responsibility, and academic integrity.

**Kewords: Generative AI; Authorship; Copyright law; Academic integrity**

## 1. Introduction

The rapid adoption of generative artificial intelligence (GenAI) tools in higher education has profoundly reshaped how students view and engage with academic tasks, particularly those involving writing, analysis, and knowledge production (Burke J et al., 2023; Kasneci et al., 2023). Software tools capable of generating coherent text, summarizing complex materials, or restructuring arguments are increasingly used by students across different fields, often as part of everyday study practices. Experimental evidence show that these tools can affect learning outcomes in different ways, depending on the context and mode of use (Deng et al., 2025). As a result, GenAI is no longer merely an external educational technology but has become an active participant in teaching and learning processes in higher education (Burke J et al., 2023).

Much of the existing debate on GenAI in higher education has focused on technical and ethical concerns such as bias, hallucinations, reliability, data protection, and privacy (Nguyen et al., 2022; Williamson & Eynon, 2020). While these issues are undisputedly important, they only partially capture the challenges posed by GenAI in academic contexts. Less attention has been paid to the nature of the interaction between humans and GenAI, particularly when such interaction directly shapes academic outputs that are traditionally understood as expressions of individual learning, reasoning, and authorship (Cotton et al., 2024; Perrotta & Selwyn, 2020).

Academic work, particularly written assignments, has long played a central role in higher education as a means of fostering cognitive development, critical thinking, and intellectual autonomy (Perrotta & Selwyn, 2020). In this context, authorship is not just a formal attribution but a substantial one, namely a proxy for responsibility, engagement, and learning achievement. However, the introduction of GenAI complicates this relationship. When students interact with GenAI systems to generate, refine, or structure academic content, the boundary between legitimate cognitive support and problematic substitution of intellectual effort becomes increasingly blurred, with its associated difficulties in assessing student progress (Eaton, 2023).

Despite the proliferation of institutional guidance on generative AI, most existing approaches regulate its use without engaging with the core legal issue raised by AI-assisted academic work, namely whether, and under what conditions, the involvement of generative systems may affect the attribution of authorship under copyright law. This gap is particularly evident in discussions of student–GenAI interaction, where recommendations often lack grounding in normative criteria capable of explaining why certain forms of interaction support learning outcomes while others undermine authorship, responsibility, and academic integrity (Cotton et al., 2024; Eaton, 2023).

We focus the present work around the legal and normative concept of authorship, and how the use of GenAI can, depending on the degree of human intervention, lead to inability of the human user to claim a work as his or hers. Indeed, several legal and institutional frameworks, notably those at the European level, offer valuable conceptual tools for addressing this issue. While copyright law and related regulatory instruments are not pedagogical theories, they are built around long-standing notions of human intellectual creation, originality, responsibility, and transparency. When translated into the context of higher education, these notions provide a structured lens through which student–GenAI interactions can be assessed, not in terms of technological capability, but in terms of their implications for learning and authorship (Nguyen et al., 2022; Williamson & Eynon, 2020).

With this work, we try to provide a conceptual and normative analysis of human–GenAI interaction in academic work, focusing on higher education settings within the European regulatory context. Instead than offering empirical measurements of learning outcomes, the article seeks to clarify the conditions under which interaction with GenAI functions as legitimate cognitive support and those under which it risks displacing authorship and undermining educational objectives (Deng et al., 2025; Kasneci et al., 2023). By integrating legal concepts of authorship and originality with pedagogical concerns about engagement and responsibility, the paper clarifies how different forms of GenAI use intersect with the requirement of human intellectual creation in academic work.

## 2. Human–GenAI Interaction, Authorship, and Responsibility in Academic Work

### 2.1. Forms of interaction with Gen-AI in academic writing

GenAI has altered not only the tools available to students, but the very forms of interaction through which academic writing is produced. Unlike earlier educational technologies, GenAI does not simply store, retrieve, or display information. Indeed, it actively produces academic texts, shaping both learning processes and outcomes (Yusuf et al., 2024).

In academic writing tasks, student interaction with GenAI may take multiple forms. These include exploratory uses such as brainstorming ideas, clarifying concepts, or generating alternative formulations, as well as more advanced forms of engagement, including drafting text, restructuring arguments, or refining language. Empirical studies suggest that such interactions can support learning when they are exercised in tandem with authentic assessment practices and require students to remain cognitively engaged with the task (Kofinas et al., 2025).

Importantly, these forms of interaction are not inherently problematic from a pedagogical standpoint. When appropriately framed, GenAI may, in some contexts, function as a cognitive scaffold, supporting learning by facilitating reflection, reducing extraneous cognitive load, or enabling students to engage more deeply with complex material. However, the educational value of these interactions depends on how they are integrated into academic work and assessment design (Selwyn et al., 2025).

*2.2. From cognitive support to cognitive substitution*

While GenAI can support learning, its use in academic writing also raises concerns about the potential substitution of students' own intellectual effort. In higher education, written academic work serves not only as an assessment tool, but also as a means of making learning processes, reasoning, and intellectual development visible. As such, the educational significance of human–AI interaction cannot be assessed solely by reference to efficiency or output quality (Williamson & Eynon, 2020).

Recent research highlights a critical distinction between *cognitive support* and *cognitive substitution*. When GenAI assists students in organizing ideas, suggesting alternative formulations, or identifying gaps in reasoning, while leaving core intellectual decisions to the human author, it may enhance engagement and learning outcomes. Conversely, when AI-generated outputs replace the students' own reasoning, analysis, or creative choices, the interaction risks undermining the educational purpose of academic work by obscuring the student's contribution (Kofinas et al., 2025; Rasul et al., 2024). In some cases, the student may be submitting a written work to which his/her intellectual contribution is scarce, which can result in not qualifying to be considered the author of the work. In such circumstances, the issue extends beyond a distortion of the learning process, as the submission may amount to a form of misrepresentation or plagiarism, thereby calling into question the validity of the assessment and, ultimately, of the academic qualification awarded.

These findings suggest that the pedagogical risks associated with generative AI lie not in its use *per se*, but in forms of interaction that displace, rather than support, student cognition (Rasul et al., 2024).

*2.3. Authorship and originality as interaction thresholds*

The concepts of authorship and originality, which will be further discussed in **Section 3**, provide a useful normative lens for analysing the boundary between cognitive support and cognitive substitution. In academic contexts, authorship is closely associated with human intellectual creation, responsibility, and the exercise of judgement. It reflects the extent to which academic work can be attributed to an individual's reasoning, decision-making, and creative choices.

From this perspective, GenAI systems cannot themselves be regarded as authors, as they lack intention, responsibility, and intellectual autonomy. Responsibility for content produced through interaction with AI therefore remains with the human user, a principle increasingly reflected in institutional policies and assessment guidelines addressing GenAI use in higher education (Gonsalves, 2025; Williamson & Eynon, 2020).

Crucially, this does not imply that any use of Gen-AI necessarily diminishes or hinders authorship. Rather, authorship functions as an *interaction threshold*: where students critically engage with AI outputs (by selecting, adapting, contextualizing, or rejecting them), the resulting work may still reflect a human intellectual contribution sufficient to ground attribution of authorship. By contrast, opaque reliance on AI-generated text, without demonstrable human engagement, weakens the link between student, authorship, and learning, calling into question both academic integrity and educational outcomes (Kofinas et al., 2025; Rasul et al., 2024; Selwyn et al., 2025).

Framing human–GenAI interaction through authorship and responsibility thus enables a nuanced evaluation of AI use in academic writing. Instead of adopting binary distinctions between permitted and prohibited uses, this approach foregrounds degrees of human involvement and intellectual engagement, aligning pedagogical objectives with emerging normative and institutional expectations in higher education (Gonsalves, 2025; Yusuf et al., 2024).

## 3. What is authorship? European Regulatory and Institutional Frameworks

As mentioned earlier, the major issue discused in this work is the boundary that separates a fair use of GenAI in opposition to a use in which human contribution is so scarce that authorship cannot be credited to the human user. Despite the much needed inputs from the fields of philosophy and pedagogy towards the concept of authorship, this discussion does not occur in a legal vacuum. Indeed, authorship is a legal concept and, in the specific case of the European level, there are several instances of binding and non-binding regulatory instruments that establish the normative principles that guide authorship, hence being directly relevant to the issue of GenAI in higher education.

Before turning to the relevant legal frameworks, a brief clarification is warranted regarding the relationship between authorship and copyright. While these concepts are closely connected, they are not synonymous. Authorship refers to the attribution of a work to a human creator whose intellectual contribution gives rise to the work itself, whereas copyright denotes the legal protection and suite of rights that follow from that attribution. In the context of this paper, copyright law is not invoked primarily as a mechanism of enforcement within academic settings, but as a normative framework that articulates the conditions under which authorship can be legally recognised. As such, copyright doctrine, notably the requirement of human intellectual creation, serves as an analytical lens for assessing when AI-assisted academic work can still be meaningfully attributed to a student as author.

### *3.1. Copyright law and the requirement of human intellectual creation*

European copyright law is premised on the notion that protection is reserved for works resulting from human intellectual creation. This principle, that in addition to several legal sources has been consistently affirmed by the Court of Justice of the European Union (CJEU), requires that a work reflect 1) the author's own intellectual choices and 2) creative freedom. Although EU copyright instruments do not yet explicitly address GenAI, this human-centric conception of authorship has significant implications for AI-assisted academic work.

At its core, this approach reflects a normative commitment to human agency. This means that copyright protection is not granted on the basis of technical sophistication, originality of output alone, or even economic value, but on the presence of a personal intellectual contribution attributable to a natural person. This conception has long functioned as a gatekeeping mechanism, distinguishing protected works from mere technical processes, mechanical reproductions, or functional outputs.

In the European space, the most relevant pieces of legislation oversseeing authorship are Directive 2001/29/EC (the "InfoSoc Directive") and Directive (EU) 2019/790 on copyright in the Digital Single Market (DSM Directive), which have been transposed into national laws with relatively high fidelity. Both directives rely and require the existence of a human author (a natural person) as the subject of rights. The boundaries of what can or cannot qualify as an original work in the context authorship have been challenged in several court cases, which decisions help to better interpret the content and reach of the law. In the specific case of the CJEU, several decisions, notably C-5/08 (Infopaq

International), C-145/10 (Painer) and C-683/17 (Cofemel) clarify that originality requires the expression of free and creative choices by a natural person. This means that outputs generated autonomously by AI systems fall outside the scope of copyright protection as "works" under EU law.

The CJEU's case law has progressively consolidated originality as a qualitative standard centred on human decision-making. In Infopaq, originality was linked to the author's "own intellectual creation"; in Painer, the Court emphasised the relevance of creative freedom and personal choices; and in Cofemel, it expressly rejected alternative standards based on aesthetic merit or technical novelty. Together, these decisions establish a consistent doctrinal position: where no (sufficient) human intellectual autonomy can be identified, there can be no authorship in the legal sense.

When applied to GenAI, this framework leads to a clear conclusion. AI systems, regardless of their level of sophistication, do not exercise creative freedom, make value-based intellectual choices, or assume responsibility for outcomes. Even where human input exists at the level of prompting, authorship depends on whether the human user retains meaningful creative control over the final result, rather than merely triggering an autonomous generative process.

In the context of higher education, this doctrinal framework reinforces the idea that GenAI cannot be regarded as an author. When students submit academic work, in order to be deemed authors of said work, with the encompassing responsibility (and benefits, such as a grade), it is necessary that the work produced results from the creative direction and intellectual contribute of a person. This reinforces the interpretation developed in **Section 2**, according to which meaningful human intellectual contribution functions as a threshold for legitimate academic authorship.

This has direct implications for assessment practices in higher education. Academic authorship operates not only as a legal concept but also as an educational proxy for learning, responsibility, and accountability. When student work is evaluated, the underlying assumption is that the submitted material reflects the student's reasoning, judgement, and intellectual development. Excessive reliance on GenAI, particularly where the system autonomously generates substantial portions of the content, risks severing this link between authorship and learning.

From this perspective, copyright law does not merely inform questions of ownership or protection, but instead provides a normative framework that helps distinguish acceptable forms of AI-assisted learning from practices that undermine the educational function of academic work. The requirement of human intellectual creation thus operates as an interaction threshold in the sense that it does not prohibit the use of GenAI, but it requires that students remain the genuine intellectual authors of the work they submit.

Accordingly, the relevance of copyright law in this context lies not in enforcing copyright claims within the classroom, but in offering a coherent and legally grounded criterion for evaluating responsibility, authorship, and originality in AI-mediated academic work.

While authorship cannot be determined in absolute or mechanistic terms, it is nevertheless possible to identify normative thresholds that are legally relevant to its attribution. To this end, we propose a qualitative assessment framework, summarised in **Figure 1**, designed to assist in evaluating authorship in GenAI-mediated academic work. Failure to satisfy any exclusionary threshold precludes attribution of authorship, whereas satisfaction of all thresholds allows authorship to be attributed, naturally subjected to contextual academic assessment.

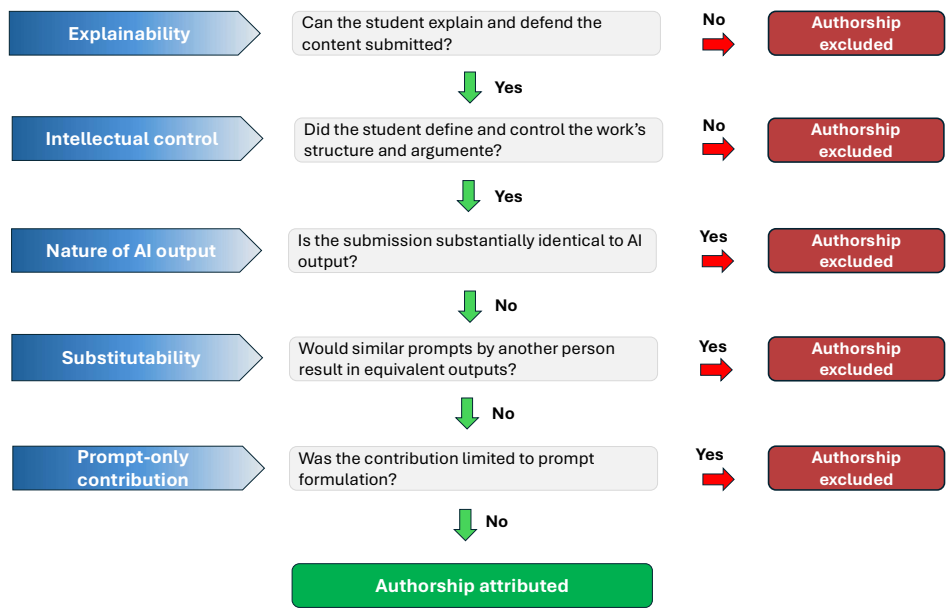

**Figure 1** - Qualitative threshold framework for assessing authorship in GenAI-assisted academic work.

### *3.2. The relevance of the AI Act: the principle of human oversight*

Regulation (EU) 2024/1689 (the "AI Act"), laying down harmonised rules on artificial intelligence, introduces a comprehensive risk-based regulatory framework for AI systems in the European Union. While general-purpose AI systems and GenAI tools are not classified as high-risk *per se*, the AI Act places particular emphasis on human oversight, transparency, and accountability.

The AI Act explicitly affirms that AI systems should support, not replace, human decision-making. Obligations relating to transparency, especially for GenAI systems capable of producing content that may be mistaken for human-generated, are particularly relevant in academic settings. These requirements resonate with higher education concerns about disclosure, responsibility, and the traceability of AI involvement in student work.

Although the AI Act does not regulate assessment or academic integrity directly, its emphasis on human control and responsibility provides a normative foundation for institutional policies governing student–AI interaction. In particular, it supports the view that responsibility for AI-assisted outputs cannot be delegated to the system itself, but remains with the human user and, where appropriate, with the institution deploying or allowinf the use of the technology.

### *3.3. Data protection, transparency, and accountability*

The General Data Protection Regulation (Regulation (EU) 2016/679 – GDPR) also plays a significant role in shaping the institutional use of GenAI in higher education. Where GenAI systems process personal data (either in training, prompting, or output generation) principles such as lawfulness, purpose limitation, transparency, and accountability become directly applicable.

From an academic perspective, transparency obligations under the GDPR reinforce the expectation that students and institutions understand how AI systems are used and what data are involved. This aligns with emerging institutional requirements for disclosure of AI use in academic work, not as a punitive mechanism, but as a means of preserving responsibility, trust, and fairness in assessment processes.

### *3.4. Institutional autonomy and emerging governance practices*

Within this European regulatory environment, universities retain significant autonomy in designing internal rules on teaching, assessment, and academic integrity. However, this autonomy operates within the boundaries set by EU law and is increasingly exercised in dialogue with suprainstitutional governance frameworks. In response to the growing use of GenAI, many European higher education institutions have adopted internal guidelines, codes of conduct, and policy documents addressing its use by students and staff, which are too numerous and qualitatively different to be addressed here.

At a suprainstitutional level, this convergence is increasingly mediated by European umbrella organisations and networks representing higher education institutions. Due to methodological limitations in assessing the position of each University towards authorship in academic works with AI use, we will focus on a brief overview of the high-level policy produced by these umbrella institutions.

The European University Association (EUA), for example, has highlighted the need for institutional approaches to AI in learning and teaching, which should be grounded in transparency, responsibility and academic integrity (EUA, 2023). Building upon these higher-level values, several Universities require, in some cases, an obligation to reference or disclose the use of AI tools in academic and student work. Indeed, it is important that such disclosure takes place, as it is pivotal for traceability and transparency, thus reinforcing the link between academic output and human responsibility. However, we find that mere disclosure of IA use is not enough, as a simple qualitative self-assessment (e.g.: I have used / not used AI) is not sufficient to assess if the student contributed with sufficient intellectual inputs to be deemed the author of said work. Indeed, we should evolve towards the disclosure, additionally, of the type of use that was given to the AI tools, for example by referring the specific sections that were contributed by AI and the level of input provided by the human user. Recently (2026), EUA has issued another document, this time advocating that the use of AI in universities must always be guided by their academic mission, values and public responsibilities (Jørgensen & Phelan, 2026).

Similarly, the European Network for Academic Integrity (ENAI) frames the ethical use of GenAI around the principles of authorized and declared use (Foltynek et al., 2023). Its recommendations emphasize that all tools influencing ideas or generating content should be properly acknowledged, including GenAI systems. While articulated primarily in the language of academic integrity, instead of copyright law, this approach strengthens a human-centred conception of authorship by ensuring that responsibility for content remains attributable to the individual student or researcher.

From the perspective of quality assurance, the European Association for Quality Assurance in Higher Education (ENQA) has issued guidelines on the responsible use of artificial intelligence that stress the primacy of human judgement and oversight (ENQA, 2025). AI systems are presented as supportive tools that must not replace human decision-making, with final evaluative responsibility remaining with academic staff. ENQA further encourages transparency regarding the use of AI, reinforcing institutional expectations of accountability in assessment and evaluation processes.

This orientation is mirrored at the level of European research governance. The European Research Area (ERA) Forum guidelines on the responsible use of GenAI in research explicitly associate honesty with the disclosure of AI use and locate accountability for outputs in human agency and oversight (ERA, 2025). Although primarily directed at research activities, these principles are highly relevant to universities as integrated academic institutions and contribute to a coherent normative environment spanning both teaching and research.

Taken together, these suprainstitutional instruments function as sector-specific soft law, shaping institutional responses to generative AI and promoting a degree of normative convergence across European higher education systems. They translate abstract regulatory principles (for example, the human-centric conception of authorship in copyright law, the requirement of human oversight under the AI Act, and the transparency and accountability obligations of the GDPR) into operational expectations for universities.

In alternative to prohibiting GenAI outright, these governance practices seek to articulate conditions under which its use is compatible with learning objectives, academic integrity, and responsibility. European regulatory and institutional frameworks thus provide a coherent multi-layered comparision matrix against which student interaction with GenAI can be evaluated: one that preserves the centrality of human authorship and responsibility while allowing space for pedagogical experimentation and institutional adaptation. These multiple layers of governace in European space are summarized in **Table 1**.

**Table 1** - Multi-level governance of generative AI in European higher education and its implications for authorship

| Level | Instrument/Entity | Legal nature | Core principle | Relevance for authorship |
|---|---|---|---|---|
| EU law | **Infosoc Directive** [Directive 2001/29/EC] **DSM Directive** [Directive 2019/790] | Binding | Human intellectual creation | Defines originality threshold |
| EU law | **AI Act** [Regulation 2024/1689] | Binding | Human oversight | Preserves human responsability |
| EU law | **GDPR** [Regulation 2016/679] | Binding | Transparency & accountability | Requires disclosure of processing |
| Suprainstitutional | EUA/ENAI/ENQA | Soft law | Disclosure; integrity | Reinforces attribution and traceability of human contribution |
| Institutional | Codes of conduct | Internal rules | Disclosure; integrity | Allocates academic responsibility and accountability to individuals |

# Conclusion

The use of GenAI in higher education challenges traditional assumptions about authorship, however it does not replace its human-centred foundations. As demonstrated, European copyright law and related regulatory frameworks remain anchored in the requirement of human intellectual creation, understood as the exercise of creative autonomy, judgement, and responsibility.

Authorship in GenAI-assisted academic work should therefore be assessed as a qualitative threshold, and not a binary condition. Where AI functions as cognitive support under meaningful human control, authorship may remain attributable to the student. Differently, where AI output substitutes human intellectual contribution, attribution becomes legally and normatively unsustainable.

By integrating copyright doctrine with principles of human oversight, transparency, and institutional governance, we argue for a context-sensitive approach that preserves the link between authorship, responsibility, and learning in AI-mediated academic work.

# List of Abbreviations

| | |
|---|---|
| **AI** | Artificial Intelligence |
| **AI Act** | Regulation EU 2024/1689 |
| **CJEU** | Court of Justice of the European Union |
| **ENAI** | European Network for Academic Integrity |
| **ENQA** | European Association for Quality Assurance in Higher Education |
| **ERA** | European Research Area |
| **EUA** | European University Association |
| **GDPR** | General Data Protection Regulation, 2016/679 |
| **Gen-AI** | Generative Artificial Intelligence |
| **Infosoc Directive** | Directive 2001/29/EC |
| **DSM Directive** | Digital Single Market Directive, 2019/790 |

# Declarations

*Availability of data and materials*

Not applicable

*Competing interests*

The author declares no competing interests

*Funding*

This work was supported by the Portuguese Foundation for Science and Technology (FCT) in the framework of Strategic Funding through the project UID/50006/2025 DOI 10.54499/UID/50006/2025 -Laboratório Associado para a Química Verde - Tecnologias e Processos Limpos.

*Authors' contributions*

The author was solely responsible for the conception, design, analysis, and writing of the manuscript. Not being a native speaker, the author has used generative AI to revise the text for clearer grammar, without affect the original intellectual contribution of the human user.

*Acknowledgements*

Not applicable

*Authors' information*

David M. Pereira holds a BSc, MSc and PhD in Pharmaceutical Sciences. He is Assistant Professor at the Faculty of Pharmacy, University of Porto, where he teaches, *inter alia*, Innovation & Intellectual Property. He also has formal legal training (BSc in Law), specializing in Technology & AI Law, as well as Intellectual Property Law.

**References**

Burke J, Crompton, H., & Burke, D. (2023). Artificial intelligence in higher education: the state of the field. *International Journal of Educational Technology in Higher Education, 20*, 22. https://doi.org/10.1186/S41239-023-00392-8

Cotton, D. R. E., Cotton, P. A., & Shipway, J. R. (2024). Chatting and cheating: Ensuring academic integrity in the era of ChatGPT. *Innovations in Education and Teaching International*, *61*(2), 228–239. https://doi.org/10.1080/14703297.2023.2190148

Deng, R., Jiang, M., Yu, X., Lu, Y., & Liu, S. (2025). Does ChatGPT enhance student learning? A systematic review and meta-analysis of experimental studies. *Computers & Education*, *227*, 105224. https://doi.org/10.1016/J.COMPEDU.2024.105224

Eaton, S. E. (2023). Postplagiarism: transdisciplinary ethics and integrity in the age of artificial intelligence and neurotechnology. *International Journal for Educational Integrity, 19*, 23. https://doi.org/10.1007/S40979-023-00144-1

European Reasearch Area (2025). *Living guidelines on the responsible use of generative AI in research*. https://research-and-innovation.ec.europa.eu/document/2b6cf7e5-36ac-41cb-aab5-0d32050143dc_en

European University Association (2023). *Artificial intelligence tools and their responsible use in higher education learning and teaching*. https://www.eua.eu/publications/positions/artificial-intelligence-tools-and-their-responsible-use-in-higher-education-learning-and-teaching.html

Foltynek, T., Bjelobaba, S., Glendinning, I., Khan, Z. R., Santos, R., Pavletic, P., & Kravjar, J. (2023). ENAI Recommendations on the ethical use of Artificial

Intelligence in Education. *International Journal for Educational Integrity*, *19*, 12. https://doi.org/10.1007/S40979-023-00133-4/METRICS

Gonsalves, C. (2025). Addressing student non-compliance in AI use declarations: implications for academic integrity and assessment in higher education. *Assessment and Evaluation in Higher Education*, *50*, 592–606. https://doi.org/10.1080/02602938.2024.2415654;WGROUP:STRING:PUBLICATION

European Association for Quality Assurance in Higher Education (2025). *Guidelines for the Responsible Use of AI in External QA • ENQA*. https://www.enqa.eu/publications/enqa-guidelines-for-the-responsible-use-of-ai-in-external-qa/

Jørgensen, T., & Phelan, C. (2026). *Adopting AI that serves the needs and values of universities*. https://www.eua.eu/publications/reports/adopting-ai-that-serves-the-needs-and-values-of-universities.html

Kasneci, E., Sessler, K., Küchemann, S., Bannert, M., Dementieva, D., Fischer, F., Gasser, U., Groh, G., Günnemann, S., Hüllermeier, E., Krusche, S., Kutyniok, G., Michaeli, T., Nerdel, C., Pfeffer, J., Poquet, O., Sailer, M., Schmidt, A., Seidel, T., … Kasneci, G. (2023). ChatGPT for good? On opportunities and challenges of large language models for education. *Learning and Individual Differences*, *103*, 102274. https://doi.org/10.1016/J.LINDIF.2023.102274

Kofinas, A. K., Tsay, C. H. H., & Pike, D. (2025). The impact of generative AI on academic integrity of authentic assessments within a higher education context. *British Journal of Educational Technology*, *56*, 2522–2549. https://doi.org/10.1111/BJET.13585

Nguyen, A., Ngo, H. N., Hong, Y., Dang, B., & Nguyen, B. P. T. (2022). Ethical principles for artificial intelligence in education. *Education and Information Technologies, 28*, 4221–4241. https://doi.org/10.1007/S10639-022-11316-W

Perrotta, C., & Selwyn, N. (2020). Deep learning goes to school: toward a relational understanding of AI in education. *Learning, Media and Technology*, *45*, 251–269. https://doi.org/10.1080/17439884.2020.1686017;SUBPAGE:STRING:FULL

Rasul, T., Nair, S., Kalendra, D., Balaji, M. S., Santini, F. de O., Ladeira, W. J., Rather, R. A., Yasin, N., Rodriguez, R. V., Kokkalis, P., Murad, M. W., & Hossain, M. U. (2024). Enhancing academic integrity among students in GenAI Era: A holistic framework. *The International Journal of Management Education*, *22*, 101041. https://doi.org/10.1016/J.IJME.2024.101041

Selwyn, N., Ljungqvist, M., & Sonesson, A. (2025). When the prompting stops: exploring teachers' work around the educational frailties of generative AI tools. *Learning, Media and Technology*, *50*, 310–323. https://doi.org/10.1080/17439884.2025.2537959

Williamson, B., & Eynon, R. (2020). Historical threads, missing links, and future directions in AI in education. *Learning, Media and Technology*, *45*, 223–235. https://doi.org/10.1080/17439884.2020.1798995

Yusuf, A., Pervin, N., & Román-González, M. (2024). Generative AI and the future of higher education: a threat to academic integrity or reformation? Evidence from multicultural perspectives. *International Journal of Educational Technology in Higher Education, 21*, 21. https://doi.org/10.1186/S41239-024-00453-6